\documentclass[twocolumn,prb,aps,amsmath,amssymb,floatfix,showpacs]{revtex4-1}

\usepackage{color}
\usepackage{graphicx}
\usepackage{dcolumn}
\usepackage{bm}
\usepackage{subfigure}

\begin{document}

\newcommand{\bo}{\boldsymbol}
\newcommand{\boq}{\mathbf{q}}
\newcommand{\bok}{\mathbf{k}}
\newcommand{\bor}{\mathbf{r}}
\newcommand{\boG}{\mathbf{G}}
\newcommand{\boR}{\mathbf{R}}

\title{Two-dimensional Fr\"ohlich interaction in 
transition-metal-dichalcogenide monolayers: 
theoretical modeling and first-principles calculations}
\author{Thibault Sohier$^{1,2}$}
\author{Matteo Calandra$^{1}$}
\author{Francesco Mauri$^{3,4}$}

\affiliation{$^{1}$IMPMC, CNRS, Universit\'e P. et M. Curie, 4 Place Jussieu, 
75005 Paris, France\\
$^{2}$ Theory and Simulation of Materials (THEOS), \'Ecole Polytechnique F\'ed\'erale de Lausanne, CH-1015 Lausanne, Switzerland\\
$^{3}$ Departimento di Fisica, Universit\`a di Roma La Sapienza, 
Piazzale Aldo Moro 5, I-00185 Roma, Italy \\
$^{4}$ Graphene Labs, Fondazione Istituto Italiano di Tecnologia
}
\date{\today}

\pacs{72.10.Di, 72.80.Ga, 73.50.Bk}

\begin{abstract}
We perform {\it ab initio} calculations of the coupling 
between electrons and small-momentum polar-optical phonons 
in monolayer transition metal dichalcogenides of the 2H type: 
MoS$_2$, MoSe$_2$, MoTe$_2$, WS$_2$, and WSe$_2$.
The polar-optical coupling with longitudinal optical phonons, 
or Fr\"ohlich interaction,
is fundamentally affected by the dimensionality of the system. 
In a plane-wave framework with periodic boundary conditions,
the Fr\"ohlich interaction is affected by the spurious interaction
between the 2D material and its periodic images.
To overcome this difficulty, we perform density functional perturbation
theory calculations with a truncated Coulomb interaction in the direction
perpendicular to the plane of the 2D material.
We show that the two-dimensional Fr\"ohlich interaction is much stronger
than assumed in previous {\it ab initio} studies.  
We provide analytical models depending on the effective charges
and dielectric properties of the materials to interpret our {\it ab initio} calculations.
Screening is shown to play a fundamental role in 
the phonon-momentum dependency of the polar-optical coupling, 
with a crossover between two regimes depending on the 
dielectric properties of the material relative to its environment.
The Fr\"ohlich interaction is screened by the dielectric environment
in the limit of small phonon momenta and sharply decreases due to 
stronger screening by the monolayer at finite momenta. 
The small-momentum regime of the {\it ab initio} Fr\"ohlich interaction is 
reproduced by a simple analytical model, for which we provide the 
necessary parameters.
At larger momenta, however, direct {\it ab initio} calculations of 
electron-phonon interactions are necessary to capture
band-specific effects.
We compute and compare the carrier relaxation times associated to 
the scattering by both LO and A$_1$ phonon modes.
While both modes are capable of relaxing carriers on timescales 
under the picosecond at room temperature, their absolute and relative
importance vary strongly depending on the material, the band, and the substrate. 
\end{abstract}

\maketitle

\section{Introduction}
Among the rapidly expanding family of two-dimensional (2D) materials, 
monolayer transition metal dichalcogenides (TMDs)
offer particularly interesting features for electronic and optoelectronic 
applications\cite{Radisavljevic2011,Wang2012a,Jariwala2014,Ganatra2014,Fiori2014}.
Thanks to high carrier mobility 
and a direct band gap in the visible range 
they can be included in 2D van der Waals heterostructures 
to fulfil various functionalities associated to 
light-matter interaction and electron transport.
In this context, it is essential to reach
a good understanding of carrier scattering\cite{Moody2015,Wang2015a,Shi2013}, 
including the intrinsic contribution from the 
electron-phonon interaction. 
In TMDs and other polar materials, a peculiar coupling emerges 
between electrons and longitudinal optical (LO) phonons. 
Such polar phonons interact with electrons by
inducing a polarization density.
At small phonon momenta, this polar-optical coupling, 
or Fr\"ohlich interaction, 
can become quite large compared to standard 
electron-phonon coupling (EPC).
Dimensionality has an interestingly drastic effect on 
this interaction. Indeed, in the limit of zero phonon momentum,
the Fr\"ohlich interaction diverges in a material with 
three-dimensional (3D) periodicity while 
it tends to a finite value in 2D materials. 
This effect can be traced back to the behaviour 
of the long-range Coulomb interaction.

Density functional perturbation theory\cite{Baroni} (DFPT) is a  
powerful tool to simulate electron-phonon interactions.
Associated to analytical models\cite{Vogl1976,Sarma1985,Mori1989},
this method can be used to establish
quantitative models\cite{Sjakste2015} of the Fr\"ohlich interaction 
in bulk materials. 
Such a comprehensive and quantitative study of the Fr\"ohlich interaction
is still missing in the case of 2D materials.
This is mainly due to the limitations of DFPT in the 2D framework.
Indeed, DFPT relies on 3D periodic boundary conditions, 
implying the presence of periodic images when simulating low-dimensional systems.
Since long-range Coulomb interactions between periodic images 
arise when low-dimensional systems are perturbed at small momenta\cite{Sohier2015}, 
DFPT fails to account for the peculiarities of the 
Fr\"ohlich interaction in 2D.
In addition to those computational limitations,
deriving analytical models of the Fr\"ohlich interaction
is not straightforward. In particular, the screening of the Coulomb 
interaction in 2D materials is a complex 
mechanism\cite{Keldysh1979,Cudazzo2011,Wehling2011,Berkelbach2013,Steinhoff2014}
requiring careful modeling.

In a previous {\it ab initio} study\cite{Kaasbjerg2012a} 
of EPC in MoS$_2$, the small-momentum behaviour of 
the 2D Fr\"ohlich interaction 
was estimated by fitting a 2D analytical model
on {\it ab initio} calculations. 
However, the calculations were performed at momenta too large 
to capture the effects of dimensionality and the analytical model only
partially accounted for the complex screening occurring 
in 2D materials.
The 2D Fr\"ohlich interaction was found to participate 
only moderately to the coupling with optical phonons in MoS$_2$, 
with a small-momentum limit three times smaller than 
the value reported here.
Consequently, it was often ignored 
in following {\it ab initio} studies of EPC
in TMDs\cite{Li2013,Jin2014}.
As far as modeling of the interaction is concerned,
a more sophisticated
model\cite{Keldysh1979,Cudazzo2011,Wehling2011,Berkelbach2013,Steinhoff2014}
of screening in 2D materials was used to estimate the strength
of the Fr\"ohlich interaction in a recent work\cite{Danovich2015}. 
This was done in the case of an isotropic dielectric tensor 
for the monolayer and without the support of direct {\it ab initio} 
computation of electron-phonon interactions.

We recently implemented\cite{Sohier2015a} the truncation of the Coulomb interaction 
between periodic images of 2D materials in the DFT and DFPT package 
Quantum ESPRESSO\cite{Giannozzi2009} (QE). 
This technique enables us to isolate each slab and
simulate electron-phonon interactions in a 2D framework.
In this work we use this approach
to compute the 2D Fr\"ohlich interaction from first principles.
We focus on the 2H polytypes of MoS$_2$, MoSe$_2$, 
MoTe$_2$, WS$_2$, and WSe$_2$.  
We propose developments on the analytical model 
of the Fr\"ohlich interaction in 2D, especially concerning its screening
in the case of a monolayer with anisotropic dielectric 
properties and for different dielectric environment 
on each side of the monolayer. 
We use {\it ab initio} calculations to estimate the parameters
of this analytical model.
The analytical model is used to interpret and support our 
calculations of the coupling to LO phonons,
and a simple effective model is proposed to reproduce 
its small-momentum limit. 
The analytical model is also used to estimate the effect of 
the presence of a substrate on the Fr\"ohlich interaction.
Finally, we compute the inverse relaxation times associated to  
intraband scattering of carriers by LO and A$_1$ phonons. 
Large variations are observed from material to material.
The relative importance of the LO and A$_1$ contributions 
strongly depends on the band in which we consider such scattering.
In any case, optical phonons (LO and/or A$_1$) are shown to be capable 
of relaxing carriers on a time scale inferior to the 
picosecond at room temperature.

\section{{\it ab initio} simulations of electron-phonon coupling}

We perform DFPT calculations of EPC in monolayer TMDs (2H type), 
using our recently developed 2D Coulomb cutoff approach\cite{Sohier2015a} within
the Quantum ESPRESSO\cite{Giannozzi2009} (QE) distribution.
This approach consists in truncating the Coulomb interaction
between the periodic images of the 2D material. 
This was implemented for the computation of total energy, forces, 
phonons and electron-phonon coupling.
The technique requires the periodic images to be separated by 
at least twice the thickness of the electronic density of the simulated layer. 
We use a separation of $\approx 17$ \AA, largely fulfilling that requirement. 
Within a slab of thickness $\approx 12$ \AA\, 
everything happens as if the monolayer was isolated.
Further details about the implementation of the 2D
Coulomb cutoff in the DFT and DFPT packages of the
QE distribution method will be exposed in a separate
publication.  
We use pseudopotentials from the Standard Solid-State Pseudopotentials (SSSP) 
library\footnote{I. E. Castelli et al. in preparation (2016), 
see http://www.materialscloud.org/sssp} (accuracy version) 
with PBE functionals and kinetic energy cutoff as indicated 
in the library. Spin-orbit coupling is neglected.
Starting from experimental lattice parameters\cite{Podberezskaya2001}, 
structures are relaxed to minimize the total energy in our DFT framework.
The resulting in-plane lattice parameters $a_0$, subsequently used in our 
calculations, are given in Table \ref{tab:param}. 
The electronic-momentum grid is set to $16 \times 16 \times 1$.
Those choices are sufficient to obtain optical phonon 
energies within a few cm$^{-1}$ of experimental values (when available).

In this section, MoS$_2$ is used as an example.
We perform calculations in bulk MoS$_2$ as well
to highlight the impact of dimensionality on the
Fr\"ohlich interaction.
For bulk MoS$_2$, we use the standard QE distribution and
the experimental\cite{Podberezskaya2001} 
out-of-plane lattice parameter of $c=12.29$ \AA. 
The corresponding unit-cell includes two layers such that 
the interlayer distance in the bulk is $\approx 6.15$ \AA.
Note that a rigorous study of the bulk requires the inclusion 
of dispersion corrections\cite{Brumme2015} to account for 
van der Waals interactions between layers.
Since we only seek a comparison of the small-momentum behaviour 
of the Fr\"ohlich interaction, however, we will ignore this aspect.

\begin{figure}[h]
\centering
\includegraphics{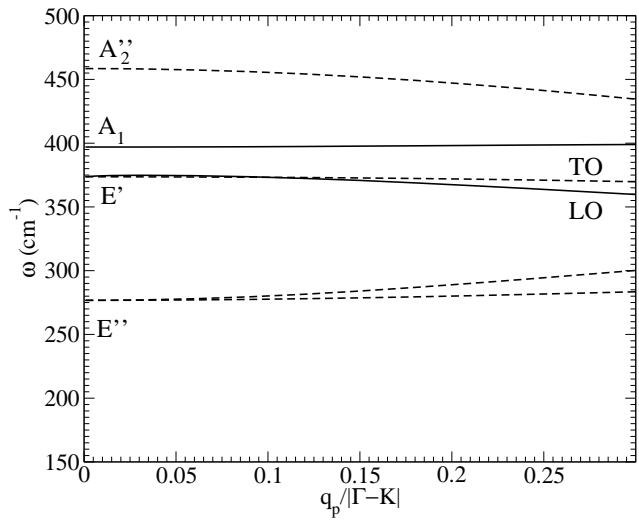}
\caption{Dispersion of the optical phonons in monolayer MoS$_2$ 
at small momenta.
For the modes in dashed lines, EPC matrix elements are negligible.
The A$_1$ and LO modes, in plain lines, couple to electrons.
We follow the notation of Ref. \onlinecite{Molina-Sanchez2011} \
for the symmetry representations of the modes at $\mathbf{\Gamma}$. 
The E' mode separates into LO and TO modes.}
\label{fig:1-OptPh}
\end{figure} 
We note $\mathbf{e}_{\boq_p \nu}$ and $\hbar\omega_{\boq_p \nu}$ the 
eigenvector and energy associated to a phonon in branch $\nu$ with in-plane 
momentum $\boq_p$. The dispersions of small-momentum optical phonons 
in MoS$_2$ are shown in Fig. \ref{fig:1-OptPh}. 
Among those, only the A$_1$ and LO modes
(plain lines in Fig. \ref{fig:1-OptPh}) couple to electrons.
In the small-momentum limit, the A$_1$ mode corresponds to
out-of-plane displacements of the sulfur atoms in phase 
opposition while the molybdenum atoms are static.
The LO mode corresponds to in-plane longitudinal displacements 
with the molybdenum atom moving in phase opposition 
to both sulfur atoms. 
A more extensive ab initio study of phonons in 
MoS$_2$ and WS$_2$ can be found in Ref. \onlinecite{Molina-Sanchez2011}.
Optical phonon modes at small momenta are qualitatively similar 
for all the TMDs studied in this work.

We consider phonon-scattering of an electron from state 
$|\bok \rangle$ to $|\bok + \boq  \rangle$ within a given band.
The associated EPC matrix element is defined as:
\begin{align}\label{eq:g2QE}
g_{\nu}(\boq_p)
&=  \sum_{a,i} \textstyle \sqrt{\frac{\hbar}{2M_a \omega_{\mathbf{q},\nu}}} \mathbf{e}^{a,i}_{\boq_p,\nu}   \langle  \mathbf{k+q} |
 \Delta^{a,i}_{\boq_p} \mathcal{V}_{\rm{KS}} (\bor) |\mathbf{k}\rangle ,
\end{align}
where $M_a$ is the mass of atom $a$ and 
$\Delta^{a,i}_{\boq_p} \mathcal{V}_{\rm{KS}} (\bor)$ is the lattice 
periodic part of the derivative of the self-consistent Kohn-Sham 
potential with respect to a phonon displacement of atom $a$ in direction $i$.

We consider neutral TMDs to avoid the metallic nature of 
the electronic screening that would occur in doped layers.
Our primary goal is the study of the long-range Fr\"ohlich interaction, 
involving LO phonons at small momenta ($|\boq_p|<15\%$ of 
$|\mathbf{\Gamma-K}|$) and an excited electron or hole.
Considering the small-momenta restriction and the energy of LO phonons, 
we can focus on intraband scattering.
We further narrow the study to the highest part of the valence band
around the high-symmetry points $\mathbf{K}$ and $\mathbf{\Gamma}$, and 
the lowest part of the conduction band around $\mathbf{K}$.
More precisely, we compute the EPC matrix elements $g_{\rm{LO}}(\boq_p)$
for the following pairs of electronic states : 
(i) $\bok=\mathbf{K}-\boq_p/2$ and $\bok+\boq_p=\mathbf{K}+\boq_p/2$ 
in the conduction band, noted "K cond" ;
(ii) $\bok=\mathbf{K}-\boq_p/2$ and 
$\bok+\boq_p=\mathbf{K}+\boq_p/2$ in the valence band, 
noted " K val" ;
(iii) $\bok=\mathbf{\Gamma}-\boq_p/2$ and 
$\bok+\boq_p=\mathbf{\Gamma}+\boq_p/2$ in the valence band, 
noted "$\Gamma$ val".
Momentum $\boq_p$ is in the $\mathbf{\Gamma} \to \mathbf{M}$ 
direction to minimize LO/TO mixing. 

\begin{figure}[h]
\centering
\includegraphics{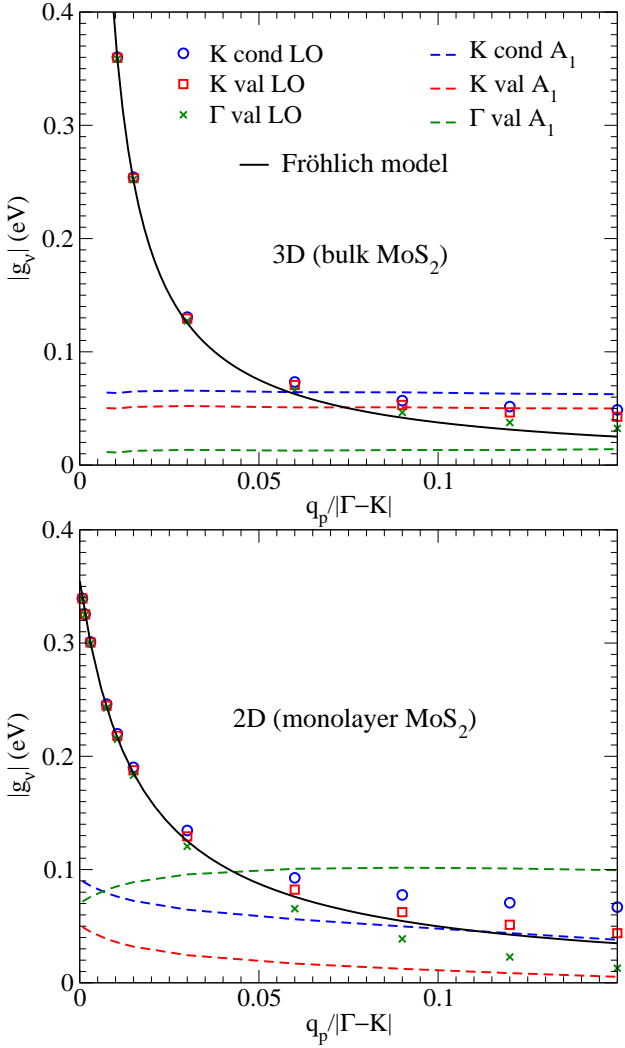}
\caption{EPC matrix elements involving LO and A$_1$ 
phonon modes in bulk and monolayer MoS$_2$. 
We consider intraband scattering of electrons or holes 
in the conduction band near $\mathbf{K}$ ("K cond") \
and in the valence band near $\mathbf{K}$ and $\mathbf{\Gamma}$
("K val" and "$\mathbf{\Gamma}$ val", respectively).
Momenta $\boq_p$ are in the $\mathbf{\Gamma} \to \mathbf{M}$ 
direction. 
The coupling to the LO phonons includes the 
Fr\"ohlich interaction. The models for 
three-dimensional and two-dimensional Fr\"ohlich 
interactions in bulk  and monolayer MoS2 
are represented in plain lines and described in the text. 
Dashed lines and symbols are DFPT calculations.
The standard QE package was used
for the bulk, while we used our implementation of 
the Coulomb cutoff in for the monolayer. }
\label{fig:2-gLO}
\end{figure}

The results of the calculations for MoS$_2$ are presented in Fig. 
\ref{fig:2-gLO}. For comparison, we add the coupling  $g_{\rm{A}_1}(\boq_p)$ 
associated to the other significant contribution of the A$_1$ mode. 
We recover the characteristic behaviours of the 
2D and 3D Fr\"ohlich interactions. 
In the 3D case, the interaction diverges as 
$ \boq_p \to \mathbf{\Gamma}$.
In the 2D framework provided by our approach, however, 
the Fr\"ohlich interaction tends to a constant at $\mathbf{\Gamma}$. 
Note that a divergence will occur when using the standard QE code, 
even if the interlayer distance is increased.
The fact that we recover the finite limit of the coupling 
at $\mathbf{\Gamma}$ thus confirms that the truncation of 
the Coulomb interaction in QE is equivalent to simulating 
an isolated monolayer.
The coupling to LO phonons at large momenta depends on the bands
via the details of the electronic wave functions.
Indeed, in that case, the variations of the 
polarization potential on the length scale of the width 
of the electronic states becomes relevant. 
Similar calculations were performed for monolayers of
MoSe$_2$, MoTe$_2$, WS$_2$, and WSe$_2$, see Fig.
\ref{fig:gtauinv} of the Appendix. 

In Figs. \ref{fig:2-gLO} and \ref{fig:gtauinv}, the plain lines 
represent analytical models discussed in the following sections.
In those models, we will focus on the more general 
small-momentum behaviour of the Fr\"ohlich interaction, 
which depends solely on the Born effective charges and dielectric 
properties of the material.
From a modeling point of view, 
the existence of finite limit at $\mathbf{\Gamma}$ for the 
2D interaction is easily established by considering the 
$1/|\boq_p|$ dependence of the 2D Coulomb interaction in reciprocal space.
The sharp decreasing of the coupling at finite-q, however, 
is a more subtle screening effect that remains to be studied in details.
Our numerical DFPT method provide us with a support to treat this issue 
in a systematic manner and establish a quantitatively accurate analytical model.

\section{Analytical Models of the Fr\"ohlich interaction}

We now present analytical models 
to explain our DFPT calculations and gain better understanding 
on the effect of dimensionality on the small-momentum 
limit of the Fr\"ohlich interaction.
The tensors of Born effective charges are noted 
$\mathcal{Z}^b_a$ and $\mathcal{Z}^m_a$ for 
bulk and monolayer, respectively. 
The index $a$ runs over the atoms of the unit cell.
The relative dielectric permittivity tensors 
(simply called dielectric tensors hereafter) 
for bulk and monolayer
are noted $\mathcal{E}^b$ and $\mathcal{E}^m$, respectively.
By symmetry, the tensors are isotropic in the plane, 
but we allow for different properties in the out-of-plane direction.
The tensors thus have the following generic forms:
\begin{align} \label{eq:tensors}
\mathcal{E}=  \begin{pmatrix}
\epsilon_p & 0 & 0\\
0 & \epsilon_p & 0\\
0 & 0 & \epsilon_z
\end{pmatrix}
\ \ \ \mathcal{Z}_a= \begin{pmatrix}
Z_{a,p} & 0 & 0\\
0 & Z_{a,p} & 0\\
0 & 0 & Z_{a,z}
\end{pmatrix} .
\end{align}
In-plane and out-of-plane variables are separated according to
the notation $\bor\to(\bor_p, z)$ and $\boq\to(\boq_p, q_z)$.
We use Gaussian CGS units.

\subsection{Three-Dimensional bulk}
We quickly recall the well-known results of the 3D case.
The small momentum behaviour of the Fr\"ohlich interaction is 
well described by the leading order in Vogl's model \cite{Vogl1976}
\begin{align}
|g^{\rm{3D}}_{\rm{Fr}} (\boq_p)|&=
\frac{4\pi e^2}{V|\boq_p|\epsilon^b_p} 
\sum_{a} \frac{ \mathbf{e}_{\boq_p} \cdot \mathcal{Z}_{a}^b 
\cdot   \mathbf{e}^{a}_{\boq_p \rm{LO}}}{\sqrt{2M_a \omega_{\boq_p\rm{LO}} } } ,
\label{eq:Fro3D}
\end{align}
where $e$ is the elementary charge, $V$ is the unit-cell's volume, 
$\epsilon^b_p$ is the in-plane dielectric constant of the bulk 
($15.37$ in MoS$_2$), and $\mathbf{e}_{\boq_p}=\boq_p/|\boq_p|$.
The pre-factor of $1/|\boq_p|$ is essentially constant in the range of 
momenta considered in this work. 
A small dependency on norm and direction of $\boq_p$ 
appears as the phonon modes deviate from the strictly longitudinal modes. 
This model is sufficient to reproduce the small-momentum
limit of the Fr\"ohlich interaction, as shown in Fig. \ref{fig:2-gLO} 
where the plain line is the above model. 

\subsection{Two-dimensional monolayer}
\label{subsec:2DModel}

\begin{figure}[h]
\centering
\includegraphics{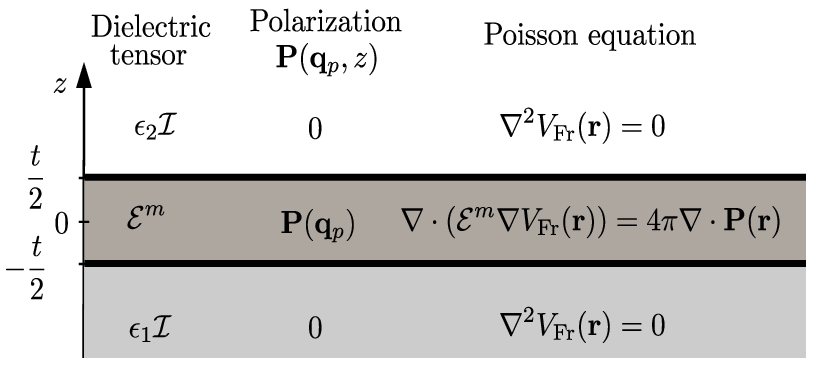}
\caption{Model of the Fr\"ohlich interaction in a polar 
2D material of thickness t. LO phonons generate a periodic 
polarization density 
$\mathbf{P}(\bor_p,z)=\mathbf{P}(\boq_p,z)e^{i\boq_p \cdot \bor_p}$ 
inside the 2D material. 
The dielectric properties of the 2D material are 
represented by the dielectric tensor $\mathcal{E}^m$
with in-plane and out-of-plane dielectric constants 
$\epsilon^m_p$ and $\epsilon^m_z$, respectively.  
Above and below are two half spaces in which the polarization is zero 
and the  dielectric constants are $\epsilon_2$ and $\epsilon_1$ respectively.
The symbol $\mathcal{I}$ denotes the identity matrix.
The two thick horizontal black lines represent surface charges at the interfaces 
of the 2D material due to the abrupt variations in the polarization density. 
We write the Poisson equation defining the Fr\"ohlich potential $V_{\rm{Fr}}$
in each region.}
\label{fig:ModelDraw}
\end{figure}

Our objective is to derive the Fr\"ohlich interaction 
in the system of Fig. \ref{fig:ModelDraw}. 
We consider LO phonons in a 2D material of thickness 
$t$. Its dielectric tensor $\mathcal{E}^m$ has the form 
of Eq. \ref{eq:tensors}, with in-plane and out-of-plane dielectric constants 
$\epsilon^m_p$ and $\epsilon^m_z$, respectively. 
Above and below are two semi-infinite spaces
with isotropic dielectric properties represented by 
the dielectric constants $\epsilon_2$ and $\epsilon_1$, respectively.

The origin of the polar-optical coupling is the polarization 
density $\mathbf{P}(\bor_p,z)$
generated by the atomic displacement pattern associated to 
a LO phonon of in-plane momentum $\boq_p$
\begin{align} \label{eq:pola}
\mathbf{P}(\bor_p,z)= \frac{e^2}{A} \sum_{a} 
\frac{\mathcal{Z}^m_{a} \cdot 
\mathbf{e}^{a}_{\boq_p \rm{LO}}}{\sqrt{2M_a \omega_{\boq_p\rm{LO}} } } f(z) 
\ e^{i \boq_p \cdot \bor_p},
\end{align}
where $A$ is the area of the unit-cell and $f(z)$ is the out-of-plane 
profile of the polarization (normalized to unity). 
Such a polarization density induces a potential
$V_{\rm{Fr}}(\bor_p, z)=
V_{\rm{Fr}}(\boq_p, z)e^{i\boq_p \cdot \bor_p}$ 
with the same periodicity. 
The associated EPC can then be written as
\begin{align}\label{eq:rhophi}
g^{\rm{2D}}_{\rm{Fr}}(\boq_p)=\int  V_{\rm{Fr}}(\boq_p, z) n_{el}(z) dz,
\end{align}
where $ n_{el}(z)$ is the plane-averaged electronic density.
By using this expression, we neglect the details of the wave-functions 
and the associated band-dependency. 
In the out-of-plane direction, we will consider the electronic density 
and the polarization to be uniform over the thickness $t$ of the material
\begin{align}
f(z)=n_{el}(z)=\frac{\theta{(t/2-|z|)}}{t},
\end{align}
where $\theta$ is the Heavyside function.
This approximation should be satisfactory in the long wavelength limit, 
since $V_{\rm{Fr}}(\boq_p, z)$ varies mildly in the out-of-plane direction.

The potential $V_{\rm{Fr}}$ must fulfil the Poisson equation
\begin{align} \label{eq:Poisson}
\nabla \cdot (\mathcal{E}(z) \nabla V_{\rm{Fr}}(\bor)) 
&=4 \pi  \nabla \cdot \mathbf{P}(\bor),
\end{align}
where $\mathcal{E}(z)$ is a position dependent 
dielectric tensor. 
The central objects of the problem are the phonon-induced 
polarization density and the dielectric tensor. 
As one travels along the out-of-plane direction, 
both those quantities change.
Inside the 2D material, $\mathcal{E}(z)=\mathcal{E}^m$ 
and the polarization density is finite and oscillating in the plane. 
Outside the 2D material, $\mathcal{E}(z)=\epsilon_{1}\mathcal{I}$ or 
$\epsilon_{2}\mathcal{I}$ (where $\mathcal{I}$ is the identity matrix)
and the polarization density is zero.
Other requirements on the potential are that that the associated 
in-plane electric field $\mathbf{E}^{\parallel}(\bor)$
and out-of-plane electric displacement $\mathbf{D}^{\perp}(\bor)$
should be continuous.

The detailed derivation of the solution to this model 
can be found in App. \ref{app:FroModel}. 
To allow for a more direct interpretation of the final solution 
in Eq. \ref{eq:FroAnaAniso}, we seek a more transparent form.
By Taylor expansion of the denominator at the linear order in 
$|\boq_p|$, the full expression of Eq. \ref{eq:FroAnaAniso}
can be recast in the form 
\begin{align}
\begin{split}\label{eq:FroEff}
| g^{\rm{2D}}_{\rm{Fr}} (\boq_p)| &= 
\frac{C_{\mathcal{Z}}}{ \epsilon_{\rm{eff}}(|\boq_p|)}
\\
\epsilon_{\rm{eff}}(|\boq_p|) &\approx \epsilon^0_{\rm{eff}}
+ r_{\rm{eff}} |\boq_p| ,
\end{split}
\end{align}
where the expressions of the parameters 
are given in table \ref{tab:expressions}.
The above form is found to reproduce Eq. \ref{eq:FroAnaAniso} 
very accurately. Only when $\epsilon^m_z \approx \epsilon_{1}$ 
or $\epsilon_{2}$ should one retain Eq. \ref{eq:FroAnaAniso} 
rather than use Eq. \ref{eq:FroEff}.
More quantitative results, depending on the nature of the monolayer, 
will be given in Sec. \ref{sec:FroEff}.
For now, let us make some qualitative remarks valid as long as 
the material is a stronger dielectric as the environment, 
which is the case of the monolayer TMDs discussed in this work, 
in vacuum or on SiO$_2$.
The bare magnitude of the polar-optical coupling is given by 
$C_{\mathcal{Z}}$.
The origin of the sharp decrease at finite
$\mathbf{q}_p$ is a screening effect specific to 2D materials. 
It can be associated with the formation
of surface charges due to the change in dielectric properties at the interfaces 
between the 2D material and its environment. 
The screening is characterized by the parameter $r_{\rm{eff}}$
which depends on the dielectric properties of the material 
as well as its thickness.
Homogeneous to a distance, it can be interpreted as an 
effective thickness marking the crossover between two screening regimes.
For $|\mathbf{q}_p| << r^{-1}_{\rm{eff}}\epsilon^0_{\rm{eff}}$, 
the coupling is screened
by $\epsilon^0_{\rm{eff}}$, which depends mainly on the dielectric 
properties of the environment. 
For $|\mathbf{q}_p| >> r^{-1}_{\rm{eff}}\epsilon^0_{\rm{eff}}$, 
the field lines are confined to the material, 
and the coupling is screened by the material. 
Materials with large dielectric constants (with respect to the environment)
will tend to focus the field lines more strongly, which results in a larger 
effective thickness $r_{\rm{eff}}$ and a sharper decrease in the 
Fr\"ohlich interaction at finite momenta (note that the slope of the 
coupling at $\mathbf{\Gamma}$ is proportional to $-r_{\rm{eff}}$).
\begin{table}[t]
\caption{Full expressions of the parameters involved in the model of the 2D Fr\"ohlich interaction, Eq. \ref{eq:FroEff}. See Fig. \ref{fig:ModelDraw}
and Eq. \ref{eq:tensors} for the definitions of the various parameters 
in the model. Note that for the isolated TMDs considered in our {\it ab initio} calculations, we have $\epsilon_1=\epsilon_2=\epsilon_{12}=1$, $\epsilon^m_z>>1$, and $\epsilon^m_p>>1$, which leads to $F\approx 1$, $\epsilon^0_{\rm{eff}}\approx 1$, and
$r_{\rm{eff}} \approx \frac{\epsilon^m_pt}{2}$.
}
{
\large
\renewcommand{\arraystretch}{2.5}
\begin{tabular}{ c @{ = } l}
\hline
$C_{\mathcal{Z}}$ &  $ \frac{2\pi e^2}{A} \sum_{a} 
\frac{\mathbf{e}_{\boq_p} \cdot \mathcal{Z}_{a}^m \cdot   \mathbf{e}^{a}
_{\boq_p \rm{LO}}}{\sqrt{2M_a \omega_{\boq_p\rm{LO}} } }  $
\\
$\epsilon^0_{\rm{eff}}$ & $ \epsilon_{12}\frac{\epsilon^m_z\bar{\epsilon}}{\epsilon^m_z\bar{\epsilon}+\epsilon_{12}(\epsilon^m_z-\bar{\epsilon})}$ 
\\
$r_{\rm{eff}}$ & $  \frac{(\epsilon^0_{\rm{eff}})^2}{\epsilon_{12}} \left(
 \frac{\epsilon_{12}}{ 3\epsilon^m_z }+ \frac{\epsilon^m_p}{2 \epsilon_{12}} F\right) \times t$ 
\\
$F$ & $ 1 + \frac{\epsilon_1\epsilon_2}{\bar{\epsilon}^2}+
\frac{\epsilon_{12}}{\bar{\epsilon}} - 
\frac{\epsilon^2_{12}}{\bar{\epsilon}\epsilon^m_z}-
\frac{\epsilon^2_{12}}{\bar{\epsilon}^2}
 -\frac{\epsilon_{12}}{\epsilon^m_z} $
\\
$\epsilon_{12}$ & $ \frac{\epsilon_1+\epsilon_2}{2}$
\ \ \ \ \ \ $\bar{\epsilon} \ =\ \sqrt{\epsilon^m_z\epsilon^m_p}$
\\
\end{tabular} }
\label{tab:expressions}
\end{table}

We have derived the general expression 
for an anisotropic slab and  different dielectric media above and below.
It can be applied to any polar material. To be quantitatively predictive, 
we need to evaluate the parameters involved.
We now detail how to evaluate those parameters 
with {\it ab initio} calculations.
 
\section{{\it ab initio} Parameters}

Here again, MoS$_2$ will be used as an example to illustrate
the method. The final parameters of interest will then be given 
for the other TMDs.
The parameters of the model are the Born effective charges, 
dielectric tensors and phonon eigenvectors.
The dynamical matrix and the corresponding phonon eigenvectors
are available from the electron-phonon calculations.
The QE code computes clamped-ions dielectric tensors 
and Born effective charges by means of linear response calculations 
with respect to an electric field perturbation\cite{Baroni}.
The Born effective charges are related to the derivative of the forces
on the atoms with respect to the applied electric field.
Since our implementation of the 2D Coulomb cutoff 
includes the computation of forces, 
the Born effective charges can be computed 
in the 2D framework for the monolayers. 
Note, however, that equivalent 
results can be obtained with the standard code. 
Indeed, Born effective charges converge relatively 
fast towards their 2D values with respect to the distance between periodic images.
The dielectric constant, on the other hand, is computed as a 
macroscopic quantity defined over a three-dimensional supercell. 
As such, the computation of the dielectric tensor of the bulk
$\mathcal{E}^b$ is straightforward and reported in 
Table \ref{tab:eps_Z} for MoS2.
The computation of an equivalent quantity 
relevant for 2D materials, however, raises issues beyond 
periodic images interactions \cite{Tobik2004,Yu2008}.
As of yet, we did not implement the modifications necessary to
compute dielectric tensors in a 2D framework.
In the following, the dielectric tensors of the monolayers
will be evaluated using the standard QE code, with an effort
to extract relevant 2D quantities from 3D calculations.  

The constant $C_{\mathcal{Z}}$ corresponds to 
the magnitude of the {\it bare} Fr\"ohlich interaction. 
It depends on the Born effective charges and the phonon displacements.
The components of the tensors $\mathcal{Z}^m_a$ (computed with 2D Coulomb cutoff)
and $\mathcal{Z}^b_a$ (computed without cutoff) 
for MoS$_2$ are given in table \ref{tab:eps_Z}.
The components of $\mathcal{Z}^m_a$ for other monolayer 
TMDs are reported in Table \ref{tab:effcharges}. 
The bare coupling $C_{\mathcal{Z}}$ varies with the direction 
and modulus of $\boq_p$ via the phonon eigenvectors.
It reaches a maximum in the $\boq_p \to \mathbf{\Gamma}$ limit, 
where the LO eigenvectors correspond to purely longitudinal modes.
It moderately decreases with increasing momenta  
($\approx - 10 \% $ at $|\boq_p|\approx 15 \% $ of $|\mathbf{\Gamma-K}|$).
Since the momentum behaviour of the Fr\"ohlich interaction 
is largely dominated by the 
screening factor $\frac{1}{\epsilon_{\rm{eff}}(\boq_p)}$, 
we can neglect the variations associated to the phonon 
eigenvectors and use the $\boq_p \to \mathbf{\Gamma}$ 
value of the bare coupling. 
Those values are reported in Table \ref{tab:param}, in the column named
"$C_{\mathcal{Z}}$ ({\it ab initio})".

\begin{table}[h]
\caption{Dielectric constants and effective charges
of bulk and monolayer MoS$_2$ as obtained in DFT.
In the case of the monolayer,
we report the dielectric constant in the case of an isotropic model.
The full range of possible values for the in-plane and out-of-plane dielectric constants is given in Fig. \ref{fig:epsoft}.}
\begin{tabular}{ c c | c c }
\hline
 \multicolumn{2}{c}{Bulk} & \multicolumn{2}{c}{Monolayer} \\
 Symbol  & Value  & Symbol  & Value  \\ 
\hline
$\epsilon^b_p$ & $15.37$ & $\epsilon^m_p=\epsilon^m_{\rm{iso}}$ & $15.5$ \\ 
 $\epsilon^b_z$ & $6.57$ & $\epsilon^m_z=\epsilon^m_{\rm{iso}}$ & $15.5$ \\  
$Z^b_{\rm{Mo},p}$ & $-0.9413$& 
$Z^m_{\rm{Mo},p}$ & $-1.0051$
\\ 
$Z^b_{\rm{Mo},z}$ & $-0.5918$& 
$Z^m_{\rm{Mo},z}$ &  $-0.0919$
\\ 
$Z^b_{\rm{S},p}$ & $0.4668$& 
$Z^m_{\rm{S},p}$ & $0.4525$
\\ 
$Z^b_{\rm{S},z}$ & $0.2921$& 
$Z^m_{\rm{S},z}$ &  $0.0411$
\\
\hline
\end{tabular}
\label{tab:eps_Z}
\end{table}
\begin{table}[h]
\caption{Effective charges of monolayer TMDs, as computed in QE 
via the response to an external electric field. 
$M \equiv$ Mo, W.$X \equiv$ S, Se, Te.}
\begin{tabular}{ c c c c c c c c c c}
\hline
Monolayer & $Z^m_{M,p}$   & $Z^m_{M,z}$ &  $Z^m_{X,p}$   & $Z^m_{X,z}$   \\ 
\hline
MoS$_2$  & -1.00 & -0.09 & 0.45 & 0.04 \\
MoSe$_2$ & -1.78 & -0.13 & 0.73 & 0.04 \\
MoTe$_2$ & -3.14 & -0.15 & 1.36 & 0.04 \\
WS$_2$   & -0.49 & -0.07 & 0.20 & 0.02 \\
WSe$_2$  & -1.17 & -0.12 & 0.43 & 0.03 \\
\hline
\end{tabular}
\label{tab:effcharges}
\end{table}

\begin{figure}[h]
\centering
\includegraphics{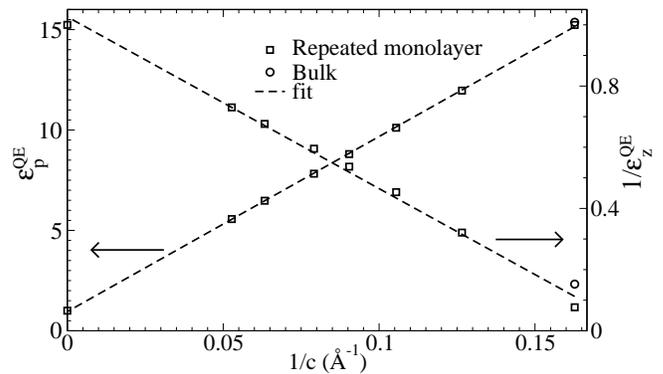}
\caption{The QE quantities $\epsilon^{QE}_p$ and $1/\epsilon^{QE}_z$ as  
functions of the inverse interlayer distance $1/c$, in MoS$_2$.
In each plot, we add the data point corresponding 
to the bulk.}
\label{fig:epsfit}
\end{figure}
\begin{figure}[h]
\centering
\includegraphics{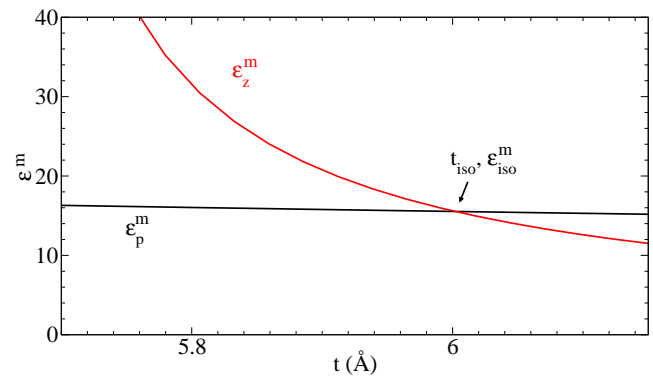}
\caption{Values of $\epsilon^m_p$ and $\epsilon^m_z$ that would be consistent 
with direct {\it ab initio} computation of the supercell's dielectric 
constants $\epsilon^{\rm{QE}}_p$ and $\epsilon^{\rm{QE}}_z$
, as functions of the corresponding thickness of the monolayer.
We indicate the point corresponding to an isotropic system.
}
\label{fig:epsoft}
\end{figure} 
We now evaluate the dielectric properties of the monolayer 
using the standard (3D) QE code. 
We simulate a system made of repeated monolayers 
separated by a varying distance $c$, with vacuum in between.
The clamped-ions dielectric tensor of this system, as computed 
within QE, is written as
\begin{align}
\mathcal{E}^{\rm{QE}} &= \begin{pmatrix}
\epsilon^{\rm{QE}}_p & 0 & 0\\
0 & \epsilon^{\rm{QE}}_p & 0\\
0 & 0 & \epsilon^{\rm{QE}}_z
\end{pmatrix}.
\end{align}
In this picture, the dielectric tensor of the bulk simply 
corresponds to $\mathcal{E}^{\rm{QE}}$ with a fixed interlayer distance $c$ 
(neglecting the small effects of an alternating stacking). 
To relate $\mathcal{E}^{\rm{QE}}$ to the dielectric tensor 
of the monolayer $\mathcal{E}^{m}$, we 
use effective medium theory and introduce 
the thickness of the monolayer as a parameter.
We then have the following relations\cite{Freysoldt2008}:
\begin{align} \label{epsQE}
\begin{split}
\epsilon^{\rm{QE}}_p &=1+(\epsilon^m_p-1)t/c \\
\frac{1}{\epsilon^{\rm{QE}}_z} &=1-\frac{(\epsilon^m_z-1)}{\epsilon^m_z}t/c.
\end{split}
\end{align}
Note that in the limit of infinite interlayer distance, 
this dielectric tensor does not tend toward $\mathcal{E}^{m}$. 
Instead, it tends towards the dielectric tensor of vacuum.

In Fig. \ref{fig:epsfit}, we plot the $\epsilon^{\rm{QE}}_p$
and $1/\epsilon^{\rm{QE}}_z$ as functions of $1/c$.
Fitting this data, we find slopes $s_1=87.2$ \AA\ 
and $s_2=5.62$ \AA\, respectively.
We then write $\epsilon^m_p$
and $\epsilon^m_z$ as functions of $t$ according to 
\begin{align}
\begin{split}
\epsilon^m_p &= 1+\frac{s_1}{t} \\
\epsilon^m_z &= \frac{t}{t-s_2} .
\end{split}
\end{align}
In principle, every set of values $\{\epsilon^m_p,\epsilon^m_z,t\}$
that satisfies the above equations can fit our DFT results.
We can assume that $t>s_2$, as we would have
$\epsilon^m_z<0$ otherwise.
We can also assume that $t<c_{\rm{bulk}} \approx 6.15$ \AA, the 
distance between two monolayers in the bulk.
In the lower panel of Fig. \ref{fig:epsfit}, we plot 
$\epsilon^m_p,\epsilon^m_z$ as functions of $t$ 
in this reasonable range of values for the thickness.
Fig. \ref{fig:epsfit} should thus be understood as a set of possible values 
for $\epsilon^m_p, \epsilon^m_z$ and the corresponding thickness.
Note that $t \approx 6$ \AA\ is consistent with the width of the 
equilibrium electronic density found in DFT.
One can see that while $\epsilon^m_p$ is almost constant
in Fig. \ref{fig:epsfit}, the variation of $\epsilon^m_z$
is more pronounced. 
Similar results are obtained for MoSe$_2$, 
MoTe$_2$, WS$_2$, and WSe$_2$. 
As far as the above ab initio study is concerned, 
we are thus left with a free parameter to model the dielectric properties
of the 2D materials, that is, a choice to make for the set of values
$\{\epsilon^m_p,\epsilon^m_z,t\}$.
For all TMDs, there is a reasonable
value of $t=t_{\rm{iso}}$ leading to an isotropic model with 
$\epsilon^m_{\rm{iso}}=\epsilon^m_p=\epsilon^m_z$.
As shown in the next section, this isotropic model is a 
choice that leads to simple yet accurate results  
for the Fr\"ohlich interaction.

\section{Effective isotropic model}
\label{sec:FroEff}

\begin{figure}[h]
\centering
\includegraphics{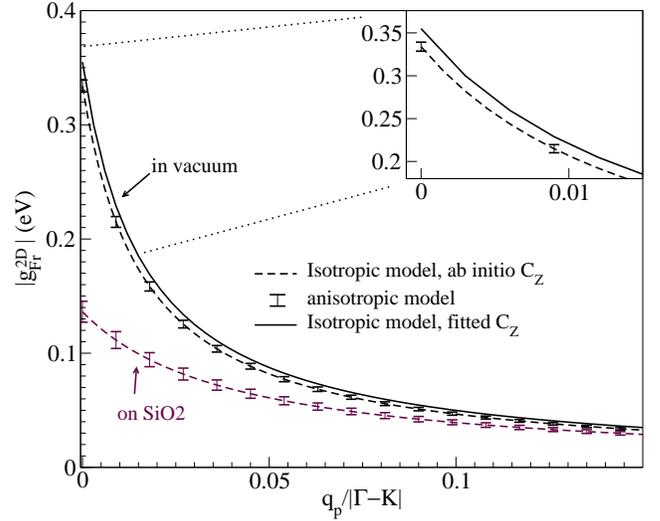}
\caption{Dashed lines are obtained by using the isotropic model 
($\epsilon^m_{\rm{iso}}=\epsilon^m_p=\epsilon^m_z$, $t=t_{\rm{iso}}$) 
and $C_{\mathcal{Z}}$ from {\it ab initio} Born effective charges.
The errors bars show the deviation from the isotropic model obtained
by using an anisotropic model with values of $\epsilon^m_p,\epsilon^m_z$
and $t$ in the range of Fig \ref{fig:epsoft}.
The plain line corresponds to the isotropic model (for MoS$_2$ in vacuum) 
with fitted $C_{\mathcal{Z}}$. This coincides with the direct DFPT calculations
of EPC, at least at small momenta (see Figs. \ref{fig:2-gLO} and \ref{fig:gtauinv}).
The inset is a zoom on the small momenta limit of the models. 
It shows that a fitted $C_{\mathcal{Z}}$ is necessary, as an anisotropic model would
not be enough to fit the direct DFPT calculations of EPC.
The SiO$_2$ substrate increases the screening of the Fr\"ohlich interaction 
strongly at small momenta.
}
\label{fig:Model}
\end{figure}
We now establish a simple effective model to reproduce 
the small-momentum limit of our direct DFPT calculations 
of the coupling to LO phonons.
We first discuss the effects of selecting different 
set of values for $\epsilon^m_p$, $\epsilon^m_z$, and $t$.
This depends on the dielectric environment, namely on the average
dielectric constant $\epsilon_{12}=\frac{\epsilon_1+\epsilon_2}{2}$.
Our DFPT calculations are performed in vacuum, with $\epsilon_{12}=1$. 
Whatever thickness we choose in Fig. \ref{fig:epsoft}, 
we have $\epsilon^m_p>>\epsilon_{12}$ and $\epsilon^m_z>>\epsilon_{12}$. 
In that case, the anisotropic model is very close to the isotropic one. 
This is shown numerically in Fig. \ref{fig:Model}. 
In dashed lines is the isotropic model, for which we use 
$\epsilon^m_{\rm{iso}}=\epsilon^m_p=\epsilon^m_z$ and $t=t_{\rm{iso}}$.
The error bars represent the deviation of the full anisotropic model 
when using other values of $\epsilon^m_p$, $\epsilon^m_z$, and $t$ within those
represented in Fig. \ref{fig:epsoft}.

\setlength{\tabcolsep}{4pt}

\begin{table*}[t]
\caption{Comparison of relevant quantities involved in the Fr\"ohlich interaction
for the monolayer TMDs MoS$_2$, MoSe$_2$, MoTe$_2$, WS$_2$, and WSe$_2$.
$a_0$ si the lattice parameter.
We report here the dielectric constants associated to the simplest isotropic model, 
that is $\epsilon^m_{\rm{iso}}=\epsilon^m_p=\epsilon^m_z$ and $t=t_{\rm{iso}}$. 
Note that we use the fact that 
$r_{\rm{eff}} \approx \epsilon^m_{\rm{iso}} t_{\rm{iso}} /2$. 
For the bare Fr\"ohlich interaction $C_{\mathcal{Z}}$, 
we report both the fitted value (from fit) used in the plots 
to reproduce the DFPT data and the value found by computing 
effective charges and phonons eigenvectors ({\it ab initio}).}
\begin{tabular}{ c c c c c c c c c c}
\hline
Monolayer & $a_0$(\AA) & $t_{\rm{iso}}$ (\AA) & $\epsilon^m_{\rm{iso}}$ 
& $r_{\rm{eff}}$ (\AA) & $C_{\mathcal{Z}}$ (eV) (from fit) &
$C_{\mathcal{Z}}$ (eV) ({\it ab initio}) 
& $\omega_{\rm{LO}}$ (cm$^{-1}$) & $\omega_{\rm{A}_1} $ (cm$^{-1}$) \\ 
\hline
MoS$_2$  & 3.18 & 6.00 & 15.5 & $46.5$ & 0.355 & 0.334 & 373.7 & 396.9 \\
MoSe$_2$ & 3.32 & 5.94 & 17.9 & $53.2$ & 0.521 & 0.502 & 277.5 & 235.4 \\
MoTe$_2$ & 3.56 & 6.65 & 20.9 & $69.5$& 0.819 & 0.819 & 223.6 & 162.9 \\
WS$_2$   & 3.18 & 5.52 & 15.2 & $42.0$ & 0.165 & 0.140 & 345.9 & 407.4 \\
WSe$_2$  & 3.31 & 5.97 & 16.3 & $48.7$ & 0.323 & 0.276 & 239.4 & 242.1 \\
\hline
\end{tabular}
\label{tab:param}
\end{table*}

For most monolayers, using the bare Fr\"ohlich interaction
$C_{\mathcal{Z}}$ calculated via the {\it ab initio} effective charges
leads to a slight mismatch with respect the direct DFPT calculations of EPC.
The effect of anisotropy in vacuum is too small to explain this mismatch, 
as seen in Fig. \ref{fig:Model}.
To reach better agreement, the parameter $C_{\mathcal{Z}}$ must be adjusted. 
The fitted values of $C_{\mathcal{Z}}$ for all monolayers are reported 
in Table \ref{tab:param}. 
Note that {\it ab initio} and fitted values stay relatively close, meaning that
a simple calculation of the effective charges can still lead to a good approximation
of the bare Fr\"ohlich interaction.
However, the mismatch is clear enough to point to some possible 
issues in the computation of the effective charges.
This imprecision on the computation of $C_{\mathcal{Z}}$ also implies that we cannot 
resolve the very small effect of anisotropy. 

Overall, an isotropic model with dielectric constant 
$\epsilon^m_{\rm{iso}}=\epsilon^m_p=\epsilon^m_z$
and a fitted $C_{\mathcal{Z}}$ (plain lines in Figs \ref{fig:2-gLO} and \ref{fig:gtauinv} of the Appendix) 
is the best choice to reproduce our DFPT results. 
Within the assumption that $\epsilon^m_{\rm{iso}}>>\epsilon_{12}$,
further simplification and greater clarity can be achieved in the model.
Indeed, the parameters of table \ref{tab:expressions} can be approximated by
\begin{align}
\epsilon^0_{\rm{eff}} &\approx \epsilon_{12}=\frac{\epsilon_1+\epsilon_2}{2} \\
r_{\rm{eff}} &\approx \frac{\epsilon^m_{\rm{iso}}}{2} t_{\rm{iso}} .
\end{align}
This simple form allows us to gain physical insight on the
screening.
In the limit $r_{\rm{eff}}|\mathbf{q}_p|>>\epsilon_{12}$, that is
$|\mathbf{q}_p|>>\frac{2\epsilon_{12}}{\epsilon^m_{\rm{iso}}t_{\rm{iso}}}$, we have 
$ g^{\rm{2D}}_{\rm{Fr}}(|\mathbf{q}_p|)
\approx g^{\rm{3D}}_{\rm{Fr}}(|\mathbf{q}_p|)$.
Indeed, the factor $1/2$ in $r_{\rm{eff}}$ enables 
to recover the prefactor of the 3D Coulomb interaction 
($2\pi e^2 \to 4\pi e^2$), while $At \approx V$, and 
$\epsilon^m_{\rm{iso}}=\epsilon^m_p\approx\epsilon^b_p$. 
In that case, the wavelength of the perturbation associated 
to the LO phonon is small and the associated field lines stay
inside the monolayer. The interaction is then screened
by the monolayer.
In the $r_{\rm{eff}}|\mathbf{q}_p|<<\epsilon_{12}$ limit, 
that is $|\mathbf{q}_p|<<\frac{2\epsilon_{12}}{\epsilon^m_{\rm{iso}}t_{\rm{iso}}}$, 
we have 
$ g^{\rm{2D}}_{\rm{Fr}}(|\mathbf{q}_p|)
\approx\frac{C_{\mathcal{Z}}}{\epsilon_{12}}$, which corresponds to the 
interaction being screened solely by the environment.
In vacuum, the crossover between those two regimes  
happens for $|\mathbf{q}_p|$ around 
$\frac{2}{\epsilon^m_{\rm{iso}}t_{\rm{iso}}}\approx 0.02$ 
\AA$^{-1}\approx 0.015 |\mathbf{\Gamma-K}|$, that is, very close to the 
$\mathbf{\Gamma}$ point. This is due to the large dielectric constant of the
monolayer compared to the environment.

An important benefit of the model is the possibility
to evaluate the effects of the dielectric environment\cite{Jena2007}.
In Fig. \ref{fig:Model}, we present results for the more experimentally
relevant case of MoS$_2$ ion SiO$_2$, for which 
$\epsilon_{12}=\frac{1+3.9}{2}=2.45$. 
The coupling is shown to be strongly decreased overall.
The validity of the approximations $\epsilon^m_p>>\epsilon_{12}$
and $\epsilon^m_z>>\epsilon_{12}$ is less clear, 
and the deviation from the isotropic model in 
Fig. \ref{fig:epsoft} is more discernible. 
However, for the purpose of estimating the effect of a SiO$_2$ substrate, 
and given the simplicity of the above parameters, 
it is still convenient to use the effective isotropic model.

The relevant parameters, including the thickness $t_{\rm{iso}}$ 
and isotropic dielectric constant $\epsilon^m_{\rm{iso}} $, 
are reported in Table \ref{tab:param} for all monolayers.
In the case of MoS$_2$, we find that the value of the coupling at 
$\mathbf{\Gamma}$ in vacuum, i.e. the bare interaction $C_{\mathcal{Z}}$, 
is three times larger than the one 
predicted in a previous {\it ab initio} study\cite{Kaasbjerg2012a}.
The bare interaction $C_{\mathcal{Z}}$ and effective screening length
$r_{\rm{eff}}$ increase with the atomic number of the chalcogen while they 
decrease with the atomic number of the transition metal.

\section{Transport}

\begin{figure*}[h]
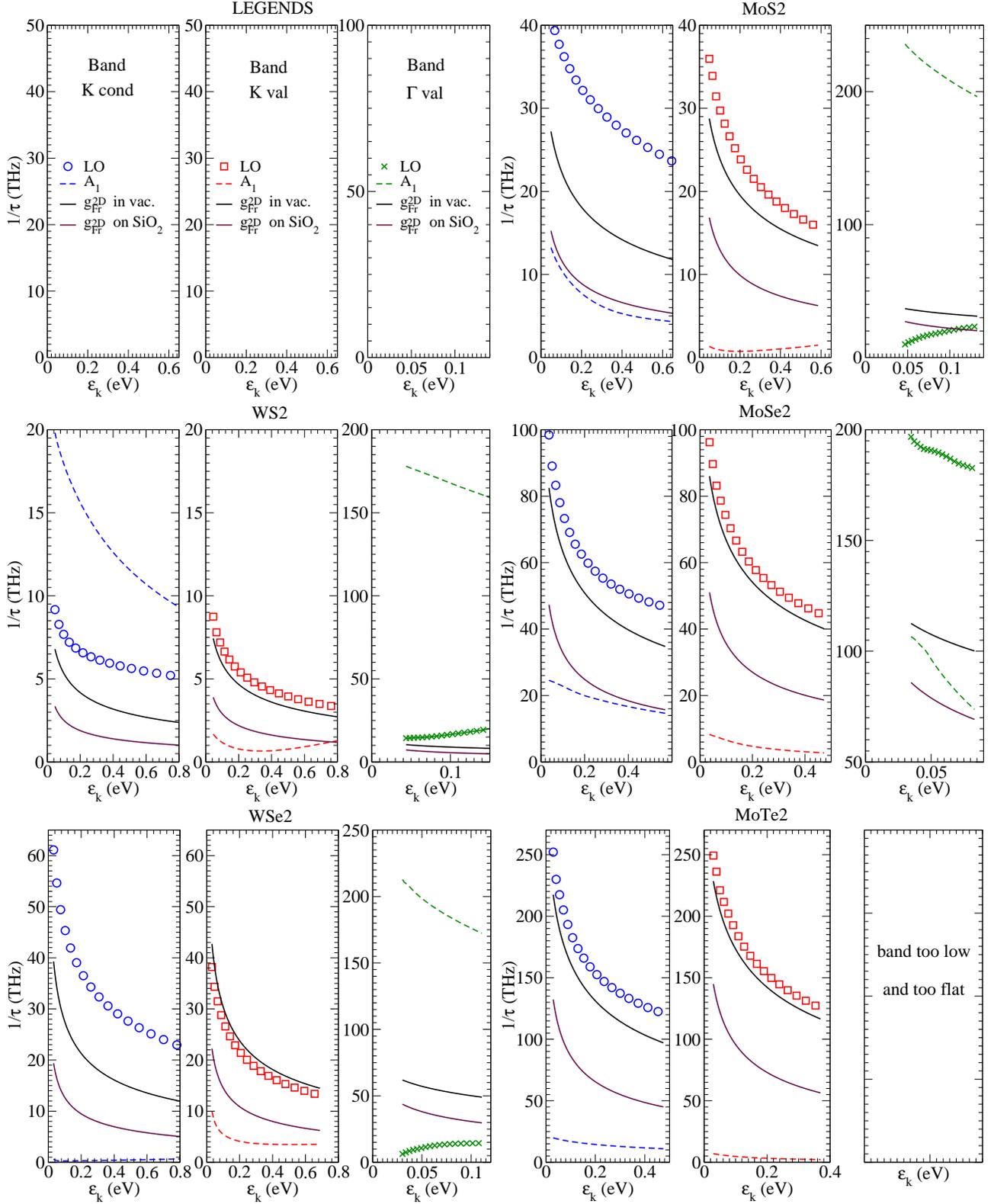

\includegraphics{7-tauinvLegends.eps}
\includegraphics{7-tauinvMoS2.eps}
\includegraphics{7-tauinvWS2.eps}
\includegraphics{7-tauinvMoSe2.eps}
\includegraphics{7-tauinvWSe2.eps}
\includegraphics{7-tauinvMoTe2.eps}
\caption{Inverse relaxation times associated to the scattering by
LO mode, A$_1$ mode and the Fr\"ohlich models, for monolayer MoS$_2$, 
MoSe$_2$, MoTe$_2$, WS$_2$, and WSe$_2$.
Calculations were performed at room temperature.
In the case of the valence band around the $\mathbf{\Gamma}$ point of MoTe2, 
the effective mass is so large that the norm of the phonon wavevectors $|\boq_p|$
involved in the scattering processes go beyond the scope of this work. In any case, 
this band is far below the valence band at $\mathbf{K}$, such that it would not be involved in transport.}
\label{fig:tauinv}
\end{figure*}

To provide a more practical sense of 
the implications of this work, we compute the following 
inverse relaxation times for an excited electron or hole 
scattered by LO or A$_1$ phonons
\begin{align} \label{eq:tauinv}
\begin{split}
\frac{1}{\tau_{\nu}(\varepsilon_{\bok})} =& \frac{2 \pi}{\hbar} 
\sum_{\boq_p} |g_{\nu}(\boq_p)|^2 \\ 
& \times \delta(\varepsilon_{\bok+\boq_p} - 
\varepsilon_{\bok} \mp \hbar \omega_{ \boq_p, \nu }) 
\begin{Bmatrix} 
N_{\nu,\boq_p}
\\
N_{\nu,\boq_p}+1
\end{Bmatrix} ,
\end{split}
\end{align}
where $\nu \equiv $LO or A$_1$, $N_{\nu,\boq_p}$ is the Bose-Einstein
distribution for phonon occupation at room temperature
and $\varepsilon_{\bok}$ is the eigenvalue energy of electronic state $|\bok\rangle$, 
measured from the bottom (top) of the conduction (valence) band.
The "$-$" (respectively "$+$") sign in the Dirac delta function $\delta$
is associated to $N_{\nu,\boq_p}$ ($N_{\nu,\boq_p}+1$) 
and corresponds to phonon absorption (emission). 
The two contributions are then summed.
In Fig. \ref{fig:tauinv}, we plot the inverse relaxation times 
for each of the three bands (K cond, K val, and $\Gamma$ val) and for each
of the MoS$_2$, MoSe$_2$, MoTe$_2$, WS$_2$, and WSe$_2$ monolayers.
To compute those quantities, we need EPC 
matrix elements on a fine grid of momenta $\boq_p$.
We use the analytical model when possible and linearly extrapolate  
the DFPT couplings otherwise.
More precisely, in the limit of small momenta, the coupling to LO phonons 
follows our analytical model of the Fr\"ohlich interaction 
and does not depend on the angle of momentum $\boq_p$ or the band.
We then use the analytical model $|g^{\rm{2D}}_{\rm{Fr}}(\boq_p)|$.
At larger momenta, the coupling depends on the band
via the wave functions. 
We then extrapolate the {\it ab initio} coupling $|g_{\rm{LO}}(\boq_p)|$.
A few other {\it ab initio} calculations were performed 
for momenta up to $|\boq^{\rm{max}}_p|\approx 0.3 |\mathbf{\Gamma-K}|$.
A mild angular dependency is possible for the {\it ab initio} matrix elements
$|g_{\rm{LO}}(\boq_p)|$ and $|g_{\rm{A_1}}(\boq_p)|$.
We neglect this angular dependency.
The integral of Eq. \ref{eq:tauinv} depends on the coupling 
and the effective masses of the corresponding band. 
We use effective masses from Ref. \onlinecite{Kormanyos2015}, reported in 
table \ref{tab:effmasses}. 
We probe electronic states with electronic momenta 
$|\mathbf{k}|<|\boq^{\rm{max}}_p|/2$. This implies that the range of
carrier energies we consider depends on the effective masses.
Note that only for MoS$_2$ is it clear that the valence band at $\Gamma$ 
should be considered. For the others, this band is lower in energy. 
For more informations about the band structures of these materials, 
see Ref. \onlinecite{Brumme2015}.
\begin{table}[h]
\caption{Effective masses are estimated using the results
of Ref. \onlinecite{Kormanyos2015}.}
\begin{tabular}{ c c c c }
\hline
Monolayer & $m^*/m_0$ $\mathbf{K}$ cond  & $m^*/m_0$ $\mathbf{K}$ val  
& $m^*/m_0$ $\mathbf{\Gamma}$ val  \\ 
\hline
MoS$_2$  & 0.45 & 0.57 & 2.52  \\
MoSe$_2$ & 0.54 & 0.65 & 3.70 \\
MoTe$_2$ & 0.56 & 0.72 & 20.0  \\
WS$_2$   & 0.31 & 0.42 & 2.17 \\
WSe$_2$  & 0.34 & 0.45 & 2.79 \\
\hline
\end{tabular}
\label{tab:effmasses}
\end{table}

Fig. \ref{fig:tauinv} shows that optical phonons are capable 
of relaxing excited carriers on a timescale inferior to the picosecond 
at room temperature.
The strength of the Fr\"ohlich interaction depends on the material considered, 
mainly via the variations of Born effective charges. 
However, this is far from being the only aspect to account 
for when studying relaxation times.
Fig. \ref{fig:tauinv} shows a great disparity of the results 
depending on the phonon mode, the band, and the material.
The analytical model of the Fr\"ohlich interaction 
is a good estimate of the DFPT calculations
only for the valence band at $\mathbf{K}$.
The relaxation times depend strongly on the band-specific,
large-momentum values of the coupling with LO phonons. 
This is due to the fact that at the minimum carrier energy 
($\varepsilon_{\bok}=\hbar\omega_{LO}$), the integral of Eq. \ref{eq:tauinv} 
already involves relatively large phonon momenta $|\boq_p|$.
The strength of the coupling with A$_1$ phonons and
thus the relative importance of the scattering by LO and A$_1$ phonons 
also depends strongly on the bands.
Very few comments apply globally. 
LO phonons seem to dominate optical-phonon scattering 
around $\mathbf{K}$, for all monolayers except WS$_2$.
A$_1$ phonons seem to dominate in the valence band around 
$\mathbf{\Gamma}$ for all monolayers except MoSe$_2$.
Although the analytical model with {\it ab initio} parameters
is useful for suspended samples in the small momentum limit 
to interpret the phenomenon, interpolate the coupling or 
to estimate the effect of the dielectric environment, 
direct DFPT calculations of EPC for each band is essential. 
The great disparity in the relaxation times and the number of
phenomenon affecting it highlight the need for 
direct {\it ab initio} simulations of electron-phonon interactions
in a two-dimensional framework. 
Furthermore, some additional effects should be included
for a quantitative comparison with experiment.
This work is a study of the coupling with optical phonons at small-momenta, 
and should provide useful guidelines to interpret 
experimental transport data. However, 
in a full quantitative study of 
transport properties, one might need to account for 
spin-orbit coupling, doping effects, the scattering of electrons in the Q band, 
the scattering between different bands... 
Those issues can be treated in the framework of QE with 2D Coulomb cutoff.

\section{Conclusion}
We have implemented the truncation of the Coulomb interaction in
the plane-wave and phonon codes of the Quantum ESPRESSO package.
This method enables us to simulate the small-momentum limit 
of the Fr\"ohlich interaction in a 2D framework, for
monolayer TMDs MoS$_2$, MoSe$_2$, MoTe$_2$, WS$_2$, and WSe$_2$.
We show that this limit is three times larger than previously 
assumed in the case of MoS$_2$ in vacuum.
We develop analytical models for the Fr\"ohlich interaction in 2D materials,
along with {\it ab initio} methods to evaluate the parameters involved.
A simple isotropic model is found to reproduce the small-momentum limit of our
DFPT calculations. We provide the parameters of this model for the various TMDs
studied.
We show that screening is paramount to evaluate the strength of the Fr\"ohlich
interaction. In particular the dielectric environment of the 2D material 
has a strong influence on the small-momentum limit of the interaction. 
Namely, the interaction is reduced by a factor
$\frac{\epsilon_1+\epsilon_2}{2}$ with respect to vacuum, 
where $\epsilon_1$ and $\epsilon_2$ are the dielectric 
constant of the environment on each side the monolayer.
We consider intraband scattering within the valence and conduction bands around $\mathbf{K}$, 
and within the valence band around $\mathbf{\Gamma}$.
Above a certain value of the momentum 
($\approx 10 \%$ of $|\mathbf{\Gamma-K}|$), the
band-dependant form of the electronic wave functions plays a 
role in the Fr\"ohlich interaction and DFPT calculations 
are necessary to evaluate deviations from the analytical model. 
Finally, we evaluate the inverse relaxation times associated to the 
scattering of photo-excited carriers by LO and A$_1$ phonons.
Those modes are shown to be capable of relaxing carriers on timescales 
smaller than the picosecond. 
The efficiency of carrier relaxation by optical phonons in TMDs is found 
to depend on many parameters.
In addition to the strength of the Fr\"ohlich interaction depending on the monolayer, 
the large momentum, band-specific coupling affects the relaxation times. 
Depending on the material and the band, 
the relaxation time associated to the A$_1$ mode 
can also be quite large. It is not correct to neglect scattering by either 
LO or A$_1$ phonons globally. 
Overall, the complexity and disparity highlighted in this work 
points to the necessity of relying on direct {\it ab initio} 
calculations of electron-phonon interactions in a 2D framework.

\section{acknowledgements}
The authors would like to thank M. Gibertini for 
his valuable help in deriving the analytical 
solution presented in appendix.
This project has received funding from the European Union’s 
Horizon 2020 research and innovation programme under 
grant agreement No 696656 GrapheneCore1
and by Agence Nationale de la Recherche under
the reference no ANR-13-IS10-0003-01. Computer facilities were
provided by CINES, IDRIS, and CEA TGCC (grant EDARI n. 2016091202).

\begin{appendix}

\section{Analytical model of the 2D Fr\"ohlich interaction}
\label{app:FroModel}
We solve here the model described in the main text, Sec. \ref{subsec:2DModel}.
The dielectric properties of the system are 
\begin{align}
\mathcal{E}(z)=
\begin{cases}
\epsilon_1 \mathcal{I} & \text{if } z < -t/2 \\
\mathcal{E}^m          & \text{if } z <|t/2| \\
\epsilon_2 \mathcal{I} & \text{if } z > t/2 
\end{cases} .
\end{align}
The potential $V_{\rm{Fr}}$ must solve the Poisson equation
\begin{align} \label{eq:Maxw}
\begin{cases}
\nabla \cdot (\mathcal{E}^m \nabla V_{\rm{Fr}}(\bor_p,z)) 
= 4 \pi \nabla \cdot \mathbf{P}(\bor_p,z) &\text{if } |z| <t/2
\\ 
\nabla^2 V_{\rm{Fr}}(\bor_p,z)
= 0 &\text{if } |z|>t/2 
\end{cases}.
\end{align}
The associated parallel electric field and orthogonal
electric displacement
\begin{align} 
\mathbf{E}^{\parallel}(\bor_p,z) 
&= - \frac{\partial V_{\rm{Fr}}(\bor_p,z)}{\partial\bor_p} , \\
\mathbf{D}^{\perp}(\bor_p,z) &= \begin{cases} 
- \epsilon_1 \frac{\partial V_{\rm{Fr}}(\bor_p,z)}{\partial z} & \text{if } z<-t/2 \\
- \epsilon^m_z \frac{\partial V_{\rm{Fr}}(\bor_p,z)}{\partial z} & \text{if } |z|<t/2 
\\
- \epsilon_2 \frac{\partial V_{\rm{Fr}}(\bor_p,z)}{\partial z} & \text{if } z>t/2
\end{cases}, \label{eq:Dperp}
\end{align}
must be continuous.

The general solution to the differential equation of Eq. \ref{eq:Maxw} is 
the sum of the solution to the homogeneous equation and a particular solution
\begin{align}
V_{\rm{Fr}}(\boq_p,z)=V_{\rm{h}}(\boq_p,z)+V_{\rm{p}}(\boq_p,z).
\end{align}
The homogeneous equation is
\begin{align} \label{eq:Maxw_homo}
\begin{cases}
\nabla \cdot (\mathcal{E}^m \nabla V_{\rm{h}}(\bor_p,z)) 
= 0 &\text{if } |z| <t/2
\\ 
\nabla^2 V_{\rm{h}}(\bor_p,z)
= 0 &\text{if } |z|>t/2 
\end{cases},
\end{align}
and the particular solution solves Eq. \ref{eq:Maxw}. 
To find a particular solution, we first solve Eq. \ref{eq:Maxw} 
inside the anisotropic material:
\begin{align}
\nabla \cdot (\mathcal{E}^m \nabla V_{\rm{p}}(\bor)) 
&=4 \pi e^2  \frac{i |\boq_p|}{A} \times \\
 & \ \  \sum_{a} 
\frac{\mathbf{e}_{\boq_p} \cdot \mathcal{Z}_{a}^m \cdot  
\mathbf{e}^{a}_{\boq_p \rm{LO}}}{\sqrt{2M_a \omega_{\boq_p\rm{LO}} } } 
f(z)  e^{i \boq_p \cdot \bor_p}  \\ 
V_{\rm{p}}(\boq_p, q_z) &= \frac{-2i C_{\mathcal{Z}} |\boq_p| }{\epsilon^m_p |\boq_p|^2+\epsilon^m_z q_z^2} f(q_z) \\
V_{\rm{p}}(\boq_p,z) &= \frac{-i  C_{\mathcal{Z}}}{\sqrt{\epsilon^m_p \epsilon^m_z} }   \int_{-\infty}^{+\infty} e^{-\tilde{\boq}_p |z'-z| } f(z') dz' 
\end{align}
with $\tilde{\boq}_p=\sqrt{\frac{\epsilon^m_p}{\epsilon^m_z}}\boq$, and $C_{\mathcal{Z}}$ is defined in table \ref{tab:expressions}. 
Using $f(z')=\frac{\theta(t/2-|z'|)}{t}$ we get for $z \in [-t/2;t/2]$
\begin{align}
V_{\rm{p}}(\boq_p,z) &=
\frac{-i C_{\mathcal{Z}}}{\sqrt{\epsilon^m_p \epsilon^m_z}} 
\frac{2}{|\tilde{\boq}_p|t}
\left( 1-e^{-|\tilde{\boq}_p| t/2 } {\rm{cosh}}(|\tilde{\boq}_p|z) 
\right),
\end{align}
where $\rm{cosh}$ is the hyperbolic cosine function.
We need to extend this particular solution outside the material.
We do not require the particular solution to carry any physical 
meaning outside the material.
It only needs to fulfil
\begin{align} 
\nabla^2 V_{\rm{p}}(\bor_p,z) 
&= 0 \ \ \  \text{if } |z|>t/2 .
\end{align}
We simply choose the solution of the above equation such that the corresponding
out-of-plane  electric field is continuous at the interfaces. 
This solution exist, and since we will only need its values at the interfaces, 
it is not necessary to specify it further.
 
Let us proceed to the homogeneous solution. Knowing that 
$V_h(\bor)=V_h(\boq_p,z)e^{i \boq_p \cdot \bor_p}$
, the homogeneous equation Eq. \ref{eq:Maxw_homo} reduces to
\begin{align} 
\begin{cases}
\frac{\partial^2 V_h(\boq_p,z)}{\partial z^2} 
&= \frac{\epsilon^m_p}{\epsilon^m_z} |\boq_p|^2 V_h(\boq_p,z) \text{ if } |z|<t/2 \\
\frac{\partial^2 V_h(\boq_p,z)}{\partial z^2} 
&= |\boq_p|^2 V_h(\boq_p,z) \text{ if } |z|>t/2
\end{cases}.
\end{align}
Adding the condition that the potential must vanish for $|z|\to \infty$, 
the solution to the homogeneous equation Eq. \ref{eq:Maxw_homo} has the form:
\begin{align}
V_{h}(\boq_p, z)=
\begin{cases}
c_3 e^{-|\boq_p|z} & \text{if } z>t/2 \\
c_1 e^{|\tilde{\boq}_p|z} +c_2 e^{-|\tilde{\boq}_p|z} & \text{if } |z|<t/2 \\ 
c_4 e^{|\boq_p|z} & \text{if } z<-t/2 
\end{cases}
\end{align}
with $\tilde{\boq}_p=\sqrt{\frac{\epsilon^m_p}{\epsilon^m_z}}\boq$.
Note that the homogeneous solution has the form of a potential 
generated by 2 surface charges at the interfaces of the monolayer.
The continuity of the potential gives
\begin{align}
c_3 e^{-|\boq_p|t/2} &= c_1 e^{|\tilde{\boq}_p|t/2} +c_2 e^{-|\tilde{\boq}_p|t/2}, \\ 
c_4 e^{-|\boq_p|t/2} &= c_1 e^{-|\tilde{\boq}_p|t/2} +c_2 e^{|\tilde{\boq}_p|t/2}.
\end{align}
The continuity of the parallel electric field is fulfilled by construction.
We use the continuity of the out-of-plane electric displacement Eq. \ref{eq:Dperp} to obtain
\begin{widetext}
\begin{align}
\begin{cases}
\frac{C_{\mathcal{Z}}}{\sqrt{\epsilon^m_p \epsilon^m_z}}
(\epsilon^m_z-\epsilon_1 ) \frac{1-e^{-|\tilde{\boq}_p|t}}{|\boq_p|t}   
 &= 
(\epsilon_1+\sqrt{\epsilon^m_p \epsilon^m_z}) c_1 e^{|\tilde{\boq}_p|t/2} 
+ (\epsilon_1-\sqrt{\epsilon^m_p \epsilon^m_z}) c_2 e^{-|\tilde{\boq}_p|t/2}
\\
\frac{C_{\mathcal{Z}}}{\sqrt{\epsilon^m_p \epsilon^m_z}}
(\epsilon^m_z-\epsilon_2 ) \frac{1-e^{-|\tilde{\boq}_p|t}}{|\boq_p|t}   
 &=
(\epsilon_2-\sqrt{\epsilon^m_p \epsilon^m_z}) c_1 e^{-|\tilde{\boq}_p|t/2} 
+ (\epsilon_2+\sqrt{\epsilon^m_p \epsilon^m_z}) c_2 e^{|\tilde{\boq}_p|t/2}
\end{cases}.
\end{align}
\end{widetext}
By defining the following dielectric mismatches
\begin{align}
&\alpha_{1}=\frac{\epsilon^m_z-\epsilon_1 }{\sqrt{\epsilon^m_p \epsilon^m_z}+\epsilon_1 } 
&\bar{\alpha}_{2}=\frac{\sqrt{\epsilon^m_p \epsilon^m_z}-\epsilon_2 }{\sqrt{\epsilon^m_p \epsilon^m_z}+\epsilon_2 } \nonumber \\
&\alpha_{2}=\frac{\epsilon^m_z-\epsilon_2 }{\sqrt{\epsilon^m_p \epsilon^m_z}+\epsilon_2 }
&\bar{\alpha}_{1}=\frac{\sqrt{\epsilon^m_p \epsilon^m_z}-\epsilon_1 }{\sqrt{\epsilon^m_p \epsilon^m_z}+\epsilon_1 } ,
\end{align}
we finally have
\begin{align}
c_1 &= \frac{C_{\mathcal{Z}}}{\sqrt{\epsilon^m_p \epsilon^m_z}} 
\frac{1-e^{-|\tilde{\boq}_p|t}}{|\boq_p|t}   
\frac{\alpha_1+ \bar{\alpha}_1 \alpha_2 e^{-|\tilde{\boq}_p|t}}{1- \bar{\alpha}_1\bar{\alpha}_2e^{-2|\tilde{\boq}_p|t}} e^{-|\tilde{\boq}_p|t/2} \\
c_2 &= \frac{C_{\mathcal{Z}}}{\sqrt{\epsilon^m_p \epsilon^m_z}} 
\frac{1-e^{-|\tilde{\boq}_p|t}}{|\boq_p|t}   
\frac{\alpha_2+ \bar{\alpha}_2 \alpha_1 e^{-|\tilde{\boq}_p|t}}
{1- \bar{\alpha}_1\bar{\alpha}_2e^{-2|\tilde{\boq}_p|t}} e^{-|\tilde{\boq}_p|t/2} .
\end{align}

The Fr\"ohlich interaction is thus
\begin{align} 
&g^{\rm{2D}}_{\rm{Fr}}(\boq_p)= \frac{1}{t}\int_{-t/2}^{t/2} \left( V_{\rm{p}}(\boq_p,z)+ V_{h}(\boq_p, z) \right) dz \\
&g^{\rm{2D}}_{\rm{Fr}}(\boq_p)=\frac{C_{\mathcal{Z}}}{\sqrt{\epsilon^m_p \epsilon^m_z}} \left[
 \frac{2}{|\tilde{\boq}_p|t} \left(  1+ \frac{e^{-|\tilde{\boq}_p|t} -1 }{|\tilde{\boq}_p|t}     \right) \right.  \label{eq:FroAnaAniso} \\
& \left. + 
\frac{(1-e^{-|\tilde{\boq}_p|t})^2}{|\boq_p|t|\tilde{\boq}_p|t} 
\frac{\alpha_1+ \alpha_2 + (\bar{\alpha}_1 \alpha_2+\bar{\alpha}_2 \alpha_1) e^{-|\tilde{\boq}_p|t}} 
{1- \bar{\alpha}_1\bar{\alpha}_2e^{-2|\tilde{\boq}_p|t}} \right].  \nonumber
\end{align}
The isotropic solution is ($\epsilon^m_p=\epsilon^m_z=\epsilon^m_{\rm{iso}}$):
\begin{align} \label{eq:FroAnaiso}
g^{\rm{2D}}_{\rm{Fr}}(\boq_p)&=\frac{C_{\mathcal{Z}}}{\epsilon^m_{\rm{iso}}} \left[
 \frac{2}{|\boq_p|t} \left(  1+ \frac{e^{-|\boq_p|t} -1 }{|\boq_p|t}     \right) \right. \\
 & \left. +
\frac{(1-e^{-|\boq_p|t})^2}{(|\boq_p|t)^2} 
\frac{\alpha_1+ \alpha_2 + 2\alpha_1 \alpha_2 e^{-|\boq_p|t}} 
{1- \alpha_1\alpha_2e^{-2|\boq_p|t}} \right] \nonumber
\end{align}

\section{Coupling with optical phonons in TMDs}

\begin{figure*}[h]
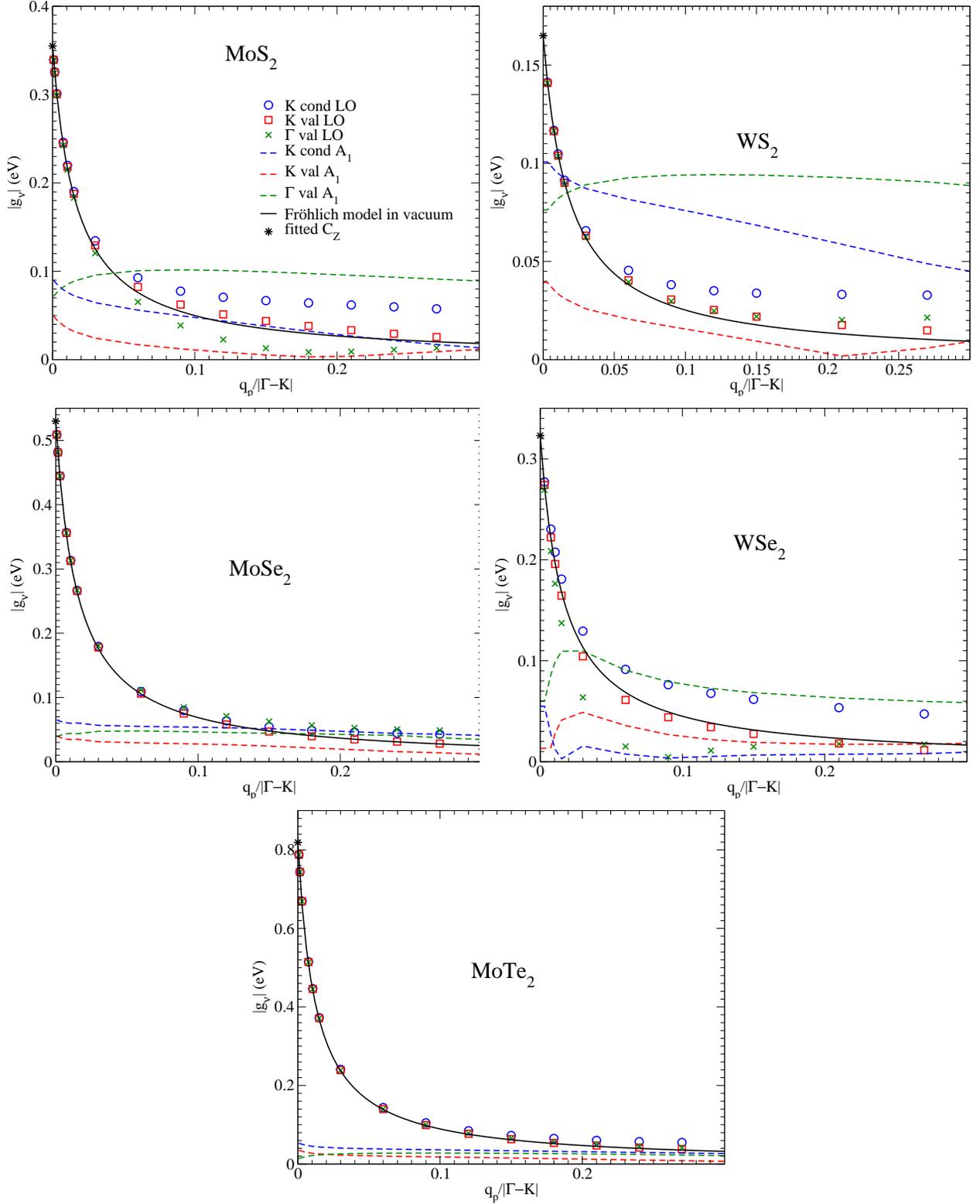

\includegraphics{gtauinvMoS2.eps}
\includegraphics{gtauinvWS2.eps}
\includegraphics{gtauinvMoSe2.eps}
\includegraphics{gtauinvWSe2.eps}
\includegraphics{gtauinvMoTe2.eps}
\caption{EPC matrix elements involving LO and A$_1$ 
phonon modes in monolayer MoS$_2$, MoSe$_2$, MoTe$_2$, WS$_2$ and WSe$_2$. 
We consider intraband scattering of electrons or holes 
in the conduction band near $\mathbf{K}$ ("K cond") \
and in the valence band near $\mathbf{K}$ and $\mathbf{\Gamma}$
("K val" and "$\mathbf{\Gamma}$ val", respectively).
Momenta $\boq_p$ are in the $\mathbf{\Gamma} \to \mathbf{M}$ 
direction. 
The analytical model of the Frohlich interaction in its simplest isotropic form and 
using the parameters indicated in Table \ref{tab:param} is shown in black plain 
lines.
Dashed lines and symbols are DFPT calculations.
}
\label{fig:gtauinv}
\end{figure*}

In Fig. \ref{fig:gtauinv} we plot the small-momentum 
coupling to the A$_1$ and LO modes
in monolayer TMDs MoS$_2$, MoSe$_2$, MoTe$_2$, WS$_2$, and WSe$_2$. 
Note that that WSe$_2$ is similar to MoS$_2$. 
WS$_2$ shows significantly smaller Frohlich interaction.
MoSe$_2$ and MoTe$_2$ are similar to each other, 
with large Fr\"ohlich interaction and some different trends in the A$_1$ mode. 
Note also that the analytical model coincides relatively well with the DFPT results
for the valence band at $\mathbf{K}$ in every material.

\end{appendix}

\bibliographystyle{apsrev4-1}
\bibliography{Frohlich}

\end{document}